\documentclass[12pt]{article}
\usepackage{graphicx,amssymb,amsmath}
\usepackage{epsfig}
\usepackage{color,graphicx}
\usepackage{cite}
\usepackage{amssymb,amsmath}
\newcommand{\be}{\begin{equation}}
\newcommand{\ee}{\end{equation}}
\newcommand{\bea}{\begin{eqnarray}}
\newcommand{\eea}{\end{eqnarray}}
\newcommand{\beas}{\begin{eqnarray*}}
\newcommand{\eeas}{\end{eqnarray*}}

\begin{document}

\begin{center}

\centerline{\Large {\bf Holographic anyonic superfluidity}}

\vspace{8mm}

\renewcommand\thefootnote{\mbox{$\fnsymbol{footnote}$}}
Niko Jokela,${}^{1}$\footnote{niko.jokela@usc.es}
Gilad Lifschytz,${}^{2}$\footnote{giladl@research.haifa.ac.il} and
Matthew Lippert${}^3$\footnote{M.S.Lippert@uva.nl}

\vspace{4mm}
${}^1${\small \sl Departamento de F\'isica de Part\'iculas} \\
{\small \sl Universidade de Santiago de Compostela}\\
{\small \sl and}\\
{\small \sl Instituto Galego de F\'isica de Altas Enerx\'ias (IGFAE)}\\
{\small \sl E-15782 Santiago de Compostela, Spain} 

\vspace{2mm}
${}^2${\small \sl Department of Mathematics and Physics} \\
{\small \sl University of Haifa at Oranim, Kiryat Tivon 36006, Israel}

\vspace{2mm}
\vskip 0.2cm
${}^3${\small \sl Institute for Theoretical Physics} \\
{\small \sl University of Amsterdam} \\
{\small \sl 1090GL Amsterdam, Netherlands} 

\end{center}

\vspace{8mm}

\setcounter{footnote}{0}
\renewcommand\thefootnote{\mbox{\arabic{footnote}}}

\begin{abstract}
\noindent
Starting with a holographic construction for a fractional quantum Hall state based on the D3-D7' system, 
we explore alternative quantization conditions for the bulk gauge fields. 
This gives a description of a quantum Hall state with various filling fractions. 
For a particular alternative quantization of the bulk gauge fields, we obtain a holographic anyon 
fluid in a vanishing background magnetic field.  We show that this system is a superfluid, exhibiting the relevant gapless excitation. 
\end{abstract}

\newpage

\section{Introduction}\label{intro}

%\subsection{Anyonic superfluid from quantum Hall state}

Anyons  are particles in $2+1$ dimensions with fractional statistics; they are neither bosonic nor fermionic.  One can model anyons as particles carrying both electric charge and magnetic flux. As two such particles circle each other, they collect an Aharonov-Bohm phase $\theta$ given by the product of the charge with the flux.
Anyons in zero magnetic field have been shown through a variety of computations to form a superfluid \cite{Fetter:1989fu,Lee:1989fw,Chen:1989xs,Lykken:1989ym} or superconductor, if they couple to a dynamical electromagnetic field. Evidence for this was given by direct computation of the relevant pole in the current-current correlator, or by exhibiting the vanishing of the Chern-Simons term in the effective field theory. 
Anyonic superfluidity is not associated with the spontaneous violation of a symmetry; instead, there is a massless mode whose existence is implied by the spontaneous violation of a fact, in this case, the fact that orthogonal translation generators commute \cite{Chen:1989xs, Giddings:1989mm}.
Alternatively, the existence of a such a massless mode can be seen by a sum rule \cite{Chen:1989xs}.

There is a close connection between the  descriptions of  the fractional quantum Hall effect and of superfluidity and superconductivity. 
Starting with a Lagrangian for a bosonic superfluid $L(\Phi)$, one can couple it to a statistical gauge field,\footnote{In this paper, statistical gauge field corresponds to a gauge field integrated in the path integral but without an $F^2$ term, while background fields are not integrated in the path integral.} add a Chern-Simons term with a fractional level, and end up with a description of the fractional quantum Hall state with particles obeying fractional statistics.\footnote{For a nice review, see\cite{Zhang:1992eu}.}  In modern language, this procedure is an $SL(2,{\mathbb Z})$ transformation.

Starting with a fluid of anyons, one can imagine taking a mean field approximation in which we consider them as fermions moving in a background of a homogeneous magnetic field induced by all the other anyons. The fermions will occupy Landau levels.  If the phase aquired by the anyons is such that $\theta=\pi(1-\frac{1}{N_L})$ for some integer $N_L$, then the fermions will be in a quantum Hall state with $N_L$ filled Landau levels. 

Quantum Hall states are gapped, incompressible, and have a filling fraction defined by 
\begin{equation}
\nu =\frac{2 \pi D}{B} \ ,
\end{equation}
where $D$ is the charge density and $B$ is the magnetic field.
The excitations are local fluctuations of the charge density.  Since the Landau levels are full, in order to make an excitation, one needs to move a charge carrier from one level to the next, at the cost of a finite amount of energy.  

However, one could imagine another kind of excitation, one which we typically do not consider when discussing a quantum Hall state.   Usually, we consider fluctuations with the magnetic field held fixed.
Instead, we can allow a perturbation where both the charge density and magnetic field fluctuate. The change in the local magnetic field will affect the degeneracy of the Landau levels and will accommodate the fluctuation of charge with no extra cost in energy provided
\begin{equation}
\delta D=\frac{\nu}{2\pi}\delta B \ .
\label{funrel}
\end{equation}
This is exactly \cite{frad, Lykken:1991xs} the gapless excitation of the anyon fluid, a signal of superfluidity. The velocity of the gapless mode for the case $\theta=\pi(1-\frac{1}{N_L})$ is \cite{Chen:1989xs} $v \sim \sqrt{N_L}\frac{\sqrt{B}}{m}$.
A similar story holds for the fractional quantum Hall state, where the anyons carry $\frac{1}{\nu}$ units of flux per unit of charge.

To be a superfluid, the massless excitations must have a linear dispersion, $\omega \sim vk$
for small $k$.  In addition, the massless mode must not be part of a continuum of states, unlike, for instance, the zero sound. One way to test whether the massless mode is discrete is to turn on a nonzero temperature.  Let us use the  intuition from the original quantum Hall state description. The quantum Hall state is gapped and is therefore unaffected by temperatures much smaller than the gap.  Because it is related to the quantum Hall state description by an $SL(2,{\mathbb Z})$ transformation, the gapless anyon fluid should be similarly robust against the effects of temperature.

\subsection{Alternative quantization}
\label{alternativequantization}

Since we want a holographic description of anyons, a natural starting point is to look at alternative quantization
 of a bulk gauge field in asymptotic $AdS_{4}$ spacetime \cite{Witten:2003ya} (see also \cite{Yee:2004ju}). The usual Dirichlet boundary condition imposed on the gauge field stems from using a bulk action which, after holographic renormalization and imposing the equation of motion, has the property that
\begin{equation}
\delta S_{D}=\int _{boundary} J_{\mu}\delta A^{\mu} \ ,
\end{equation}
where $J_{\mu}=\frac{\delta S_{D}}{\delta A^{\mu}}$ is a conserved current interpreted as the conserved current in the CFT. Consistency of the variation principle or, equivalently, requiring no flux through the time-like boundary at infinity requires imposing the condition that $A_{\mu}$ is fixed at the boundary. However, this is not the only consistent boundary condition one can impose on a gauge field in $AdS_{4}$. Both independent solutions of the linear  equation of motion for the bulk gauge field are normalizable, and one can consider the theory with other boundary conditions, which again guarantee that no information is lost through time-like infinity \cite{Ishibashi:2004wx,Marolf:2006nd }.  For example, one can start with a new bulk action 
\begin{equation}
S_{N}=S_{D}-\int _{boundary} J_{\mu}A^{\mu} \ ,
\end{equation}
where $N$ stands for Neumann boundary conditions.  The variation now gives
\begin{equation}
\delta S_{N}=-\int_{boundary}A^{\mu}\delta J_{\mu} \ .
\end{equation}
Now we see that with this action, one instead needs to fix $J_{\mu}$ at the boundary. 
To make things more symmetric, and for later formalism, it is convenient to introduce a new quantity $v_{\nu}$: Since
$\partial^{\mu}J_{\mu}=0$, one can write $J_{\mu}=\frac{1}{2\pi}\epsilon_{\mu \rho \nu}\partial^{\rho}v^{\nu}$. Therefore, $v_\nu$ is 
defined only up to gauge transformations.
The variation of the action is now
\begin{equation}
\delta S_{N}=-\frac{1}{2\pi}\int_{boundary}\epsilon _{\mu \rho \nu}\partial^{\rho} A^{\nu}\delta v^{\mu}\equiv- \int_{boundary}J^{*}_{\mu}\delta v^{\mu} \ ,
\end{equation}
where the new current  $J^{*}_{\mu}=\frac{1}{2\pi}\epsilon_{\mu \rho \nu}\partial^{\rho}A^{\nu} \equiv {\cal B}_{\mu}$.

In the CFT, the Neumann boundary condition is related to the Dirichlet boundary condition by an $S$ operation, giving a new CFT which includes a gauge field.
One can now consider more general modified boundary terms for the bulk action. 
Since $J_{\mu}A^{\mu}=\frac{1}{2\pi}\epsilon_{\mu \rho \nu}\partial^{\rho}v^{\nu}A^{\mu}$ we can consider adding more  general such terms with  $v_{\mu}$ and $A_{\mu}$ that are invariant under gauge transformations:
\begin{equation}
v_{\mu}, A_{\mu} \rightarrow v_{\mu}+\partial_{\mu} V\quad , \ \  A_{\mu} +\partial_{\mu} \Lambda \ .
\end{equation}
Thus, we can write
\begin{equation}
S_{gen}=S_{D}+\frac{1}{2\pi}\int_{boundary}[a_1\epsilon_{\mu \rho \nu}A^{\mu}\partial^{\rho}v^{\nu}+a_2\epsilon_{\mu \rho \nu}A^{\mu}\partial^{\rho}A^{\nu}+a_3\epsilon_{\mu \rho \nu}v^{\mu}\partial^{\rho}v^{\nu}] \ .
\end{equation}
Now the variation of the action takes the form
\begin{equation}
\delta S_{gen}=\int_{boundary} (a_{s}J_{\mu}+b_{s} {\cal B}_{\mu})(c_{s}\delta v^{\mu}+d_{s} \delta A^{\mu}) \ ,
\end{equation}
where
\bea
a_sd_s=1+a_1, \ \ b_sc_s=a_1,\ \ b_s d_s=2a_2, \ \ a_s c_s=2a_3 \ .
\eea
Evidently, $a_{s}d_{s}-b_{s}c_{s}=1$ and so 
\be
\left(\begin{array}{cc}a_s & b_s \\c_s & d_s\end{array}\right)
\ee
forms an $SL(2,{\mathbb R})$ matrix.
The new boundary condition holds $c_{s} v^{\mu}+d_{s} A^{\mu}$ fixed, or in gauge invariant form,
\begin{equation}
{{\cal B}}^{*}_{\mu}=c_{s} J_{\mu}+d_{s}  {\cal B}_{\mu}= {\rm fixed} \ \  \rightarrow \ \ \delta{{\cal B}}^{*}_{\mu}=0 \ .
\label{newcon}
\end{equation}
The new current is just
\begin{equation}
{J}^{*}_{\mu}=a_sJ_{\mu}+b_s {\cal B}_{\mu} \ .
\label{newj}
\end{equation}

Since 
\begin{equation}
J_{\mu} \sim F_{\mu u}\equiv {\cal E}_{\mu} \ ,
\end{equation}
where $u$ is an inverse radial coordinate with the boundary at $u=0$, we see that this $SL(2,{\mathbb R})$ action is an electric-magnetic duality transformation. We do not mean that the bulk action has to be invariant under this transformation.  One only needs that the equation of motion for the bulk gauge field reduces near the boundary to the free equation, so that one can choose mixed boundary conditions.

If the bulk theory has only integer electric and magnetic charges, as in string theory, for consistency the $SL(2,{\mathbb R})$ transformations must be  restricted to a subset $SL(2,{\mathbb Z})$, acting on approprietly normalized quantities. 

From the point of view of the dual CFT, the $SL(2,{\mathbb Z})$ transformations are certain operators changing one CFT to another \cite{Witten:2003ya}. For example, starting with a Lagrangian $\tilde{L}(\Phi,A)$ for the fields of the CFT  $\Phi$ coupled to an external vector field $A$,  the $S$ operation, for which $a_s=d_s=0$ and $b_s=1=-c_s$, yields the same Lagrangian but now with $A$ as a dynamical field. The new CFT also has a conserved $U(1)$ current $\frac{1}{2\pi}\epsilon_{\mu \rho \nu}\partial^{\rho}A^{\nu}$,
and one can couple to an external vector field $C$ by
$\frac{1}{2\pi}\epsilon_{\mu \rho \nu}C^{\mu}\partial^{\rho}A^{\nu}$. So, the pair that transforms under $SL(2,{\mathbb Z})$ are $J_{\mu}$ and $ {\cal B}_{\mu}$. The new boundary condition can be read from the $A_{\mu}$ equation of motion.

The $T$ operation, where $a_s=b_s=d_s=1$ and $c_s=0$, just adds a Chern-Simons term for the external  vector field (say $C_{\mu}$),
\be
\frac{1}{2\pi}\epsilon_{\mu \rho \nu}C^{\mu}\partial^{\rho}C^{\nu}.
\ee 
The $SL(2,{\mathbb Z})$ transformations are generated by $S$ and $T$ transformations, which do not commute with each other.

From equation (\ref{newcon}), we see that the new current is a current of particles carrying (with respect to the original definitions) $c_s/d_s$ units of original magnetic flux for each unit of original charge. These are anyons.

In this paper, we consider a holographic model of a fractional quantum Hall state and impose alternative boundary conditions appropriate to change it into an anyonic fluid in zero magnetic field. We then show that this fluid is a superfluid by analyzing the fluctuations and exhibiting a gapless mode in the current-current correlation function. Since the original fermions were strongly coupled through an $SU(N)$ gauge field, the resulting anyons should also be considered as strongly coupled.

%%%%%%%%%%%%%%%%%%%%%%%%%%%%%%%%%%%%%%%%%%%%%%%%
\section{Holographic model}

We will begin with the D3-D7' model of the fractional quantum Hall effect \cite{Bergman:2010gm}.\footnote{For an earlier  use of a closely related  system, see \cite{Rey:2008zz}, and for other holographic approaches to the quantum Hall effect, see \cite{otherqh}.  In addition, the D2-D8' model \cite{Jokela:2011eb, Jokela:2011sw,Jokela:2012se}, a close cousin of the D3-D7' model, exhibits analogous properties to those presented here, including anyonic superfluidity.}  This model has a rich phenomenology, with both a gapped quantum  Hall phase and a ungapped metallic phase \cite{Bergman:2011rf, Jokela:2012vn, Davis:2011gi,Semenoff}.  We will focus here on the gapped quantum Hall phase, whose fluctuations were analyzed in \cite{Jokela:2010nu}.

%%%%%%%%%%%%%%%
\subsection{D7-brane action}
The model consists of a probe D7-brane in the background of $N$ D3-branes, oriented such that their intersection is $(2+1)$-dimensional.  
The metric of the D3-brane background is
\be 
 ds_{10}^2 = \frac{r^2}{L^2}(-h(r)dt^2+dx^2+dy^2+dz^2)+\frac{L^2}{r^2}\left(\frac{dr^2}{h(r)}+r^2 d\Omega_5^2\right) \ ,
\ee 
where $h(r)=1-\frac{r_T^4}{r^4}$ and
\be
 d\Omega_5^2 = d\psi^2 + \cos^2\psi(d\theta^2+\sin^2\theta d\phi^2)+\sin^2\psi(d\alpha^2+\sin^2\alpha d\beta^2) \ .
\ee
The AdS radius is given by $L^2=\sqrt{4\pi g_{s} N}\alpha'=\sqrt{\lambda}\alpha'$, and the temperature is $T = \frac{r_T}{\pi L^2}$.
In addition, the RR four-form $C^{(4)}_{txyz}=-\frac{r^4}{L^4}$.  The angles have the following 
ranges: $\psi \in [0, \pi/2]$, $\theta$, $\alpha \in [0, \pi]$, and $\phi$, $\beta \in [0, 2\pi]$.

We will consider embeddings of the D7-brane which span the $t$, $x$, $y$, and $r$ directions and wrap the two $S^2$'s.
For the fluctuation analysis, it is convenient to introduce the Cartesian coordinates \cite{Jokela:2010nu}
\bea
 \rho & = & r \sin\psi \\
 R    & = & r \cos\psi \ .
\eea
We take $R$ as the worldvolume coordinate, while $\rho(R)$ and $z(R)$ describe the embedding.  For stability, we turn on a flux through one $S^2$ \cite{Myers:2008me, Bergman:2010gm}
\be
2\pi\alpha' F_{\alpha\beta} = \frac{f L^2}{2}\sin\alpha \ .
\ee
With flux only on one $S^2$, a constant $z$ is a solution to the equations of motion.  We can thus set it to zero in what follows. The induced metric on the D7-brane is then
\bea
 ds^2_{D7} &=& \frac{r^2}{L^2} \left(-h \ dt^2+dx^2+dy^2\right)
 +\frac{L^2R^2}{r^2}d\Omega_2^{(1)2}+\frac{L^2\rho^2}{r^2}d\Omega_2^{(2)2} \nonumber\\
 &&+\frac{L^2}{r^2} \left(\frac{R^2}{hr^2}+\frac{\rho^2}{r^2}-2\left(\frac{1}{h}-1\right)\frac{\rho R}{r^2}\rho'+ \left(\frac{\rho^2}{hr^2} + \frac{R^2}{r^2}\right)\rho'^2\right)dR^2 \ ,
\eea
where primes denote derivatives with respect to $R$.  To include a background charge density and magnetic field, we turn on:
\be
 2\pi\alpha' F_{xy} = b \ , \ 2\pi\alpha' F_{Rt} =  a'_0 \ , \ 
 \ee
Finally, we perform the following rescaling  of the coordinates and gauge fields:
\be\label{eq:rescaling}
 R = L\sqrt b \tilde R \ , \ \rho = L\sqrt b \tilde\rho \ , \ r = L\sqrt b \tilde r\ , \ a_0 = L\sqrt b \tilde a_0 \ ,
\ee
which scales out $b$ and $L$.

The D7-brane action in these coordinates is
\be
\label{action}
 S = -{\cal N}\int d\tilde R\left(\tilde R^2\sqrt{f^2+4\frac{\tilde\rho^4}{\tilde r^4}}\sqrt{1+\tilde r^{-4}}\sqrt A-c(\tilde R)\tilde a'_0  \right) \ ,
\ee
where ${\mathcal N} = 8\pi^2 T_7 V_3 L^5 b^{3/2}$ and
\be
 \tilde r^2 A = \tilde R^2-\tilde r^2\tilde a'^2_0-2(h-1)\tilde R\tilde\rho\tilde\rho'+h\tilde R^2\tilde\rho'^2+\tilde\rho^2(h+\tilde\rho'^2) \ .
\ee
The pull-back of the RR-potential onto the worldvolume, $c(\tilde{R})$, is given by
\be
 c(\tilde{R}) = \arctan(\tilde{\rho}/\tilde{R})-\frac{1}{4}\sin(4\arctan(\tilde{\rho}/\tilde{R}))-\psi_\infty+\frac{1}{4}\sin(4\psi_\infty) \ ,
\ee
where
\be
\psi_\infty =\lim_{\tilde{R} \rightarrow \infty}\arctan \frac{\tilde{\rho}}{\tilde{R}} \ ,
\ee
which is related to the flux $f$ by
\be
f^2=4\sin^{2} \psi_{\infty}-8\sin^{4} \psi_{\infty} \ .
\ee

%%%%%%%%%%%%%%%%%%%%%%%%%
\subsection{Background equations of motion}

The equation of motion for the gauge field, integrated once, gives
\bea
\frac{\tilde g}{h}(1+\tilde r^{-4})\tilde a'_0 & = & \frac{d}{b}-2c(\tilde R)\equiv \frac{\tilde d}{b}  \ ,
\eea
where
\bea
 \tilde g & = & \frac{2h\tilde R^2\sqrt{f^2+4\frac{\tilde\rho^4}{\tilde r^4}}}{\sqrt{1+\tilde r^{-4}}\sqrt A} 
\eea
and $d$ is proportional to the charge density $D$. 
These can be solved:
\bea
 \tilde a'_0 & = & \frac{\tilde d}{b}\frac{h}{\tilde g(1+\tilde r^{-4})} = \frac{\tilde d}{b}\sqrt{\frac{\tilde r^{-2}\left( \tilde R^2-2(h-1)\tilde R\tilde\rho\tilde\rho'+h\tilde R^2\tilde\rho'^2+\tilde\rho^2(h+\tilde\rho'^2) \right)   }{\left(\frac{\tilde d}{b}\right)^2+4\tilde R^4(1+\tilde r^{-4})\left(f^2+4\frac{\tilde \rho^4}{\tilde r^4}\right)}} \nonumber\\
 \tilde g & = & \frac{h}{1+\tilde r^{-4}}\sqrt{\frac{ \left(\frac{\tilde d}{b}\right)^2+4\tilde R^4(1+\tilde r^{-4})\left(f^2+4\frac{\tilde \rho^4}{\tilde r^4}\right)}{    \tilde r^{-2}\left( \tilde R^2-2(h-1)\tilde R\tilde\rho\tilde\rho'+h\tilde R^2\tilde\rho'^2+\tilde\rho^2(h+\tilde\rho'^2) \right)         }} \ .
\eea
The embedding scalar equation of motion is
\bea
 & & \partial_{\tilde R}\left(\frac{(1+\tilde r^{-4})\tilde g}{h\tilde r^2}\left\{(h\tilde R^2+\tilde\rho^2)\tilde\rho'+(1-h)\tilde R\tilde\rho  \right\}\right) \\
  & = & -2\tilde a'_0\partial_{\tilde\rho}c(\tilde R)+\frac{32h\tilde R^6\tilde\rho^3}{\tilde g\tilde r^6}-\frac{8h\tilde R^4\rho\left(f^2+4\frac{\tilde\rho^4}{\tilde r^4}\right)}{\tilde g\tilde r^6(1+\tilde r^{-4})}+\frac{\tilde g}{2h}(1+\tilde r^{-4})\partial_{\tilde\rho}A \nonumber \ ,                     
\eea
where
\bea
 \partial_{\tilde\rho}c(\tilde R) & = & \frac{8\tilde R^3\tilde\rho^2}{\tilde r^6} \\
 \partial_{\tilde\rho}A & = & \frac{2}{\tilde r^4}(1-h)(\tilde R\tilde\rho'-\tilde\rho)(\tilde R^2-2\tilde\rho^2+3\tilde R\tilde\rho\tilde\rho'). \nonumber 
 \eea

\subsection{Minkowski embeddings}

In this system there are two types of embeddings:  black hole (BH) embeddings, where the D7-brane enters the horizon at $\tilde r=\tilde r_T$, and Minkowski (MN) embeddings, where the D7-brane caps off at some $\tilde r_0>\tilde r_{T}$.   We will focus here exclusively on MN embeddings.  

The D7-brane can end smoothly if the tip is located at $\psi = \pi/2$ ({\emph{i.e.}}, $\tilde R = 0$) where the $S^2$ without wrapped flux degenerates.  The tip can not support any local charge density, so a MN embedding requires $\tilde{d}=0$, implying
\begin{equation}
\label{MNfillingfraction}
\frac{d}{b}=2c(0)=\pi-2\psi_{\infty}+\frac{1}{2}\sin (4\psi_{\infty}) \equiv  \frac{\pi \nu}{N} \ .
\end{equation}
Here $\nu=\frac{2\pi D}{B}$ is the filling fraction, and $D$ and $B$ are the physical charge density and magnetic field. From (\ref{MNfillingfraction}), we see that $\nu$ is fixed by the choice of $\psi_{\infty}$, or actually of $f$.

\begin{figure}[ht]
\center
\includegraphics[width=0.8\textwidth]{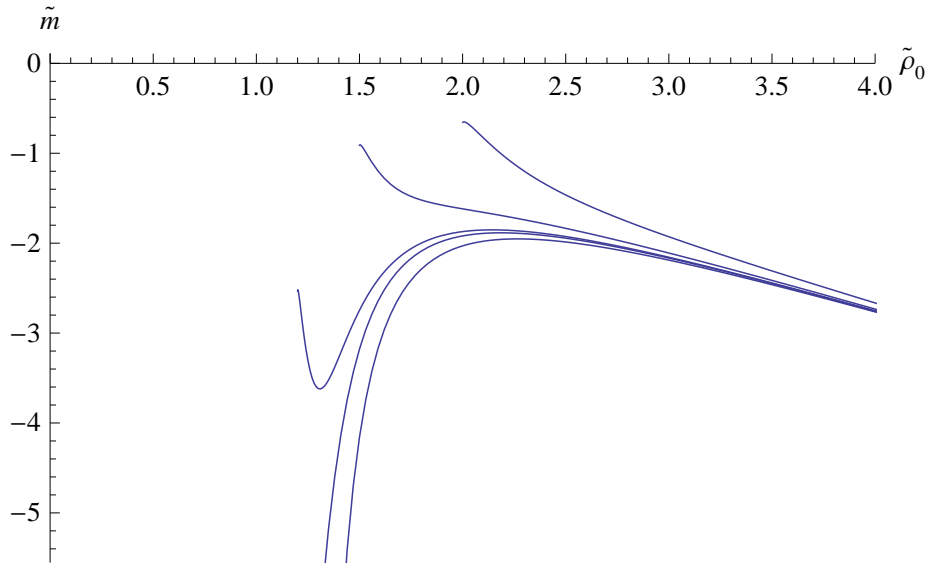}
\caption{$\tilde{m}$ vs $\tilde{\rho}_{0}$ for MN embeddings, at various temperatures:  from top to bottom, $\tilde r_T = 2$, 1.5, 1.2, 1.1, and 0.}
\label{fig:mvsrho0}
\end{figure}

For a fixed $\psi_{\infty}$, the leading UV behavior of an embedding is characterized by 
\be
\tilde m = \tilde r^{-\Delta_{+}}\sin\left(\arctan\left(\frac{\tilde\rho}{\tilde R}\right)-\psi_\infty\right) \ ,
\ee
where
\be
\Delta_{\pm}=-\frac{3}{2}\pm \frac{1}{2}\sqrt{73-\frac{48}{\cos^{2} \psi_{\infty}}} \ .
\ee
The dimensionless parameter $\tilde m$ is proportional to the mass $m$ of the fermions in the dual boundary theory:
\be
\tilde{m}=\frac{2\pi \alpha'}{L}\frac{m}{b^{-\Delta_+/2}} \ .
\ee
The relationship between $\tilde{m}$ and $\tilde{\rho}_{0}$ for various temperatures is given in Fig.~\ref{fig:mvsrho0}.

Because the D7-brane avoids the horizon, there is a gap for charged fluctuations.  The size of the gap is given by the length of a string stretched from the horizon to the D7-brane tip, which is proportional to $\tilde\rho_0\equiv \tilde\rho(\tilde R=0)$.  The neutral fluctuations, analyzed in \cite{Jokela:2010nu}, are also gapped.  

The conductivities for the MN phase were computed in \cite{Bergman:2010gm}.  The longitudinal conductivity is
\begin{equation}
\label{sigmaxx}
\sigma_{xx}=0 
\end{equation}
which is a consequence of the mass gap.  The Hall conductivity is
\be
\sigma_{yx}=\frac{D}{B}=\frac{\nu}{2\pi}=\frac{2Nc(0)}{\pi}
\label{qhs}
\ee
as expected for a quantum Hall fluid.  Note that, in string theory, the charge density  is the number of strings per unit volume and the magnetic field is $2 \pi$ times the $D5$-brane number per unit volume.  Therefore, the filling fraction $\nu=\frac{2\pi D}{B}$ is always rational.

We can view this MN state in two different ways.  In one description, we have a theory with $N$ species of quark, and the basic excitation is a quark with charge one and mass $m$.  Each species is then in a fractional quantum Hall state with filling fraction $\frac{\nu}{N}$. Alternatively, we can view the basic excitation as a baryon made out of $N$ quarks, carrying charge $N$ and having mass of order $N$.  Then, the filling fraction for the baryon is just $\nu$.  In the bulk description, the baryons are represented by smeared D5-branes, which, due to the background of a RR 3-form flux, source a charge on the D7-brane.  The quarks are then the zero-length strings stretching from the D5-branes to D7-brane.

\subsection{Alternative quantization}

The application of $SL(2,{\mathbb Z})$ (or some subgroup of it) to quantum Hall systems has a long history,\footnote{See \cite{Rey:1989jg, Burgess:1999ug, Burgess:2000kj, Burgess:2006fw} and references therein.} and recently $SL(2,{\mathbb Z})$ duality has been incorporated in holographic quantum Hall models \cite{Goldstein:2010aw, Bayntun:2010nx}.  To understand the role of $SL(2,{\mathbb Z})$, let us start with a quantum Hall state, with the conductivity given by (\ref{sigmaxx}) and (\ref{qhs}) and gapped to neutral excitations.   If we turn on a constant background electric field in the $x$-direction $E_x$, we obtain a current in the $y$-direction
\begin{equation}
J_{y}=\frac{D}{B}E_{x}.
\label{hcur}
\end{equation}
We perform the $SL(2,{\mathbb Z})$ transformation on the current\footnote{The components of the current are $J_{\mu}=(-D,\vec{J}_{i})$, and also $2\pi{\cal B}_{\mu}=(B,E_{y},-E_{x})$.}
 and the background electromagnetic field given by (\ref{newcon}) and (\ref{newj}):
\begin{eqnarray}
&{J}^{*}_{x}=0, \hspace{2.3cm} &{E}^{*}_{y}=0 \nonumber\\
&{J}^{*}_{y}=a_s J_{y} -b_s \frac{E_{x}}{2\pi}, \hspace{0.5cm} &-\frac{{E}^{*}_{x}}{2\pi}=c_s J_{y} -d_s \frac{E_{x}}{2\pi} \nonumber\\
&-{D}^{*}=-a_s D +b_s \frac{B}{2\pi}, \hspace{0.5cm} &\frac{{B}^{*}}{2\pi}=-c_s D +d_s \frac{B}{2\pi} \ .
\label{jetrans}
\end{eqnarray}
Under this transformation, the filling fraction transforms\footnote{In some conventions the filling fraction transforms under  $SL(2,{\mathbb Z})$ with opposite signs for $c_s$ and $b_s$, which is also an $SL(2,{\mathbb Z})$ transformation.}
\begin{equation}
\label{nuSL2Z}
\nu^{*}  = \frac{a_s\nu-b_s}{-c_s\nu +d_s}
\end{equation}
and similarly for $\sigma_{yx}$.
Thus, we can start with some filling fraction $\nu$ which is fixed by the embedding, in particular, by $\psi_{\infty}$, and an $SL(2,{\mathbb Z})$ transformation will yield a theory with a different conserved current $ {J}^{*}_\mu$ and a different  filling fraction $\nu^*$. The gravity dual for the new theory has the same bulk action and thus the same equations of motion and the same embedding.  The only difference is in the boundary terms, as explained in Sec.~\ref{alternativequantization}.  Consequently, the statistics of the excitations change under $SL(2,{\mathbb Z})$ as well.

Starting with the Hall current (\ref{hcur}), if we choose $c_s$ and $d_s$ such that $c_sD-d_s\frac{B}{2\pi}=0$, we get that both ${E}^{*}_{x}=0$ and ${B}^{*}=0$ while ${J}^{*}_{y} \not= 0$.
We end up with a nonvanishing current but without a background electric field driving it and without a background magnetic field.\footnote{Note that this is not possible to achieve if $\sigma_{xx} \neq0$.}   Hence, this is a candidate for an anyonic  superfluid.  Note that the original quantum Hall state broke parity and time reversal invariance\footnote{See \cite{Davis:2011gi} for how discrete symmetries act on the bulk solution.} due to the external magnetic field.  The anyonic superfluid therefore also breaks parity and time reversal but without any external magnetic field.

\section{Fluctuation analysis}

We now compute the spectrum of collective excitations around the candidate anyon superfluid.
We analyze the fluctuations of all fields around the MN background as in \cite{Jokela:2010nu} but impose alternative boundary conditions on the bulk gauge fields.  If this system really is a superfluid, we expect to see a gapless excitation.

%fluctuations
We work in radial gauge ($a_{\tilde R} = 0$) and consider fluctuations with the following wavelike form:
\bea
\delta \tilde a_\mu &=&  \delta \tilde a_\mu(\tilde R) e^{-i\omega t + ikx} \\
\delta \tilde \rho &=&  \delta \tilde \rho(\tilde R) e^{-i\omega t + ikx} \ .
\eea
Fluctuations of the embedding coordinate $z$ decouple completely, so we will ignore them here.  In order to completely scale out the magnetic field, we need to rescale the frequency and momentum by
\be
(\omega,k) = \frac{\sqrt{b}}{L} (\tilde\omega, \tilde k) \ .
\ee
It is useful to define the gauge-invariant combination
\be
\delta e_x =  \omega\delta a_1 + k\delta a_0
\ee
which rescales as $\delta \tilde e_x = \delta e_x/b$.  Expanding the D7-brane action (\ref{action}) to quadratic order, we derive the fluctuation equations of motion.  This system of coupled, linear equations can be found in Appendix~\ref{appendix}.

%alternative BC
The boundary conditions for the fluctuations of the D7-brane worldvolume gauge field using a general alternative quantization condition (\ref{newcon}) can be put in the convenient form
\be
\label{alternativeboundarycondition}
0 = -n \delta F_{\mu u} + \frac{1}{2} \epsilon_{\mu\nu\rho} \delta F^{\nu\rho} \ ,
\ee
where the indices $\mu, \nu$, and $\rho$ are $(2+1)$-dimensional boundary coordinates, raised and lowered by the flat metric $\eta_{\mu\nu}$, and the inverse radial coordinate $u = L^2/r$.  

 We rewrite the relationship (\ref{funrel}) in the rescaled variables:
\be
\delta d =(\pi-2\psi_{\infty}+\frac{1}{2}\sin (4\psi_{\infty})) \delta b \ .
\ee 
Comparing this with (\ref{alternativeboundarycondition}) and noting that
\be
F_{\mu u} (u=0)=\frac{d}{[4\cos^4 \psi_{\infty}(f^2+4\sin^4 \psi_{\infty})]^{1/2}} \ ,
\ee
we find the parameter $n$ is given in terms of the parameters of the $SL(2,{\mathbb Z})$ transformation (\ref{nuSL2Z}) by
\be
\label{ndef}
 n = \frac{N}{\pi [4\cos^4 \psi_{\infty}(f^2+4\sin^4 \psi_{\infty})]^{1/2}}\frac{c_s}{d_s} \ .
\ee
In components, the boundary condition (\ref{alternativeboundarycondition}) is
\bea
n \delta \partial_u a_0 + ik\delta a_2 & = & 0
\label{mu=0}\\
 n \delta \partial_u a_1 - i\omega\delta a_2 & = & 0
\label{mu=1}\\
 -in \delta \partial_u a_2 + \delta e_x & = & 0 \ ,
\label{mu=2}
\eea
where we have used $\delta e_x =  i\omega\delta a_1 + i k\delta a_0$.
We can now use the gauge constraint coming from the equation of motion for $\delta a_z$, which, for $u \rightarrow 0$, reads 
\be
k \delta \partial_u a_1 + \omega\delta\partial_u a_0 = 0 \ .
\ee
We can then show that both (\ref{mu=0}) and  (\ref{mu=1}) end up giving the same boundary condition:
\be
\label{mu=0or1}
0 = \frac{in}{\omega^2 - k^2} \delta \partial_u e_x + \delta a_2 \ .
\ee

Now we need to put the boundary conditions (\ref{mu=0or1}) and (\ref{mu=2}) in terms of the rescaled radial coordinate $\tilde R$, which we will use in the numerical calculations. 
In terms of $\tilde R$ and the other rescaled variables, the boundary conditions are
\bea
\frac{-in}{\tilde \omega^2 - \tilde k^2}\frac{\tilde R^2}{\cos\psi_\infty} \delta \partial_{\tilde R} \tilde e_x + \delta\tilde a_2 & = & 0 \nonumber\\
  i n \frac{\tilde R^2}{\cos\psi_\infty}  \delta \partial_{\tilde R} \tilde a_2 + \delta \tilde e_x & = & 0 \ .
\label{bcfinal}
\eea

From equations (\ref{newcon}) and (\ref{newj}), we see  that if we vary the action twice with respect to $c_s\delta v^{\mu}+d_s \delta A^{\mu}$, we get the two-point function of the new current.  If we vary with respect to $c_s\delta J^{\mu}+d_s \delta {\cal B}^{\mu}$, we get the two-point function of $a_s v^{\mu}+b_s A^{\mu}$, and then acting with the operator $\epsilon_{\nu \rho \mu}\partial^{\rho}$, we also get the two-point function of the new current. %I don't understand the point of this paragraph, should it maybe appear somewhere else?

We search for normal modes by looking for pairs $(\tilde{\omega}, \tilde{k})$ for which there is a solution to the fluctuation equations (\ref{atEOM}), 
(\ref{axEOM})-%, \ref{ayEOM}, \ref{rhoEOM}, \ref{zEOM},
(\ref{aREOM}) 
with the boundary conditions (\ref{bcfinal}).
We use the fluctuation analysis techniques used in \cite{Jokela:2010nu} and based on \cite{Kaminski:2009dh, Amado:2009ts}.
As explained in Sec.~\ref{intro}, we expect that there will be a gapless mode for a certain $n_{crit}$,
\be
n_{crit}=\frac{[4\cos^4 \psi_{\infty}(f^2+4\sin^4 \psi_{\infty})]^{1/2}}{\pi-2\psi_{\infty}+\frac{1}{2}\sin (4\psi_{\infty})} \ .
\label{ncomp}
\ee
Of course, whether $n_{crit}$ can be realized by an $SL(2,{\mathbb Z})$ transformation (rather than an $SL(2,{\mathbb R})$ transformation) depends sensitively on $\psi_{\infty}$ and on the number of D3-branes $N$, so we will not worry about this.

\subsection{Numerical results}
\begin{figure}[ht]
\center
\includegraphics[width=0.475\textwidth]{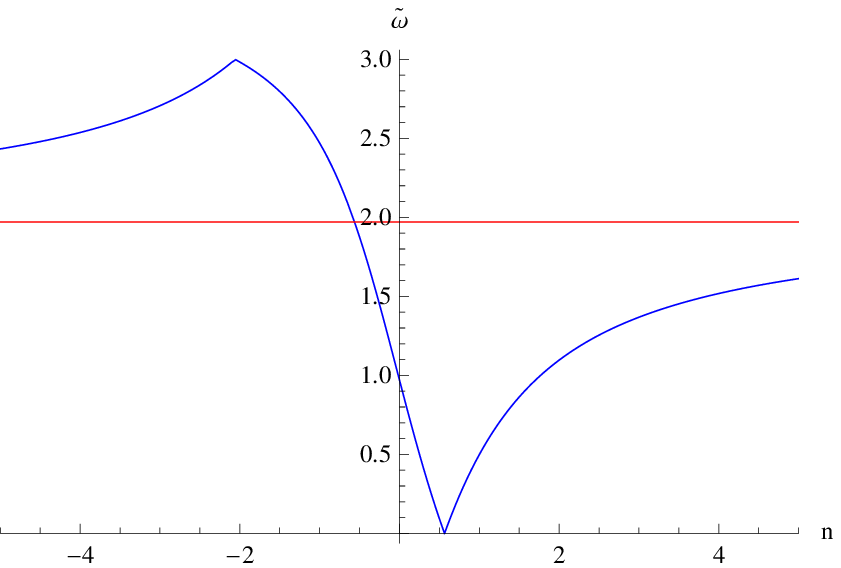}
\includegraphics[width=0.475\textwidth]{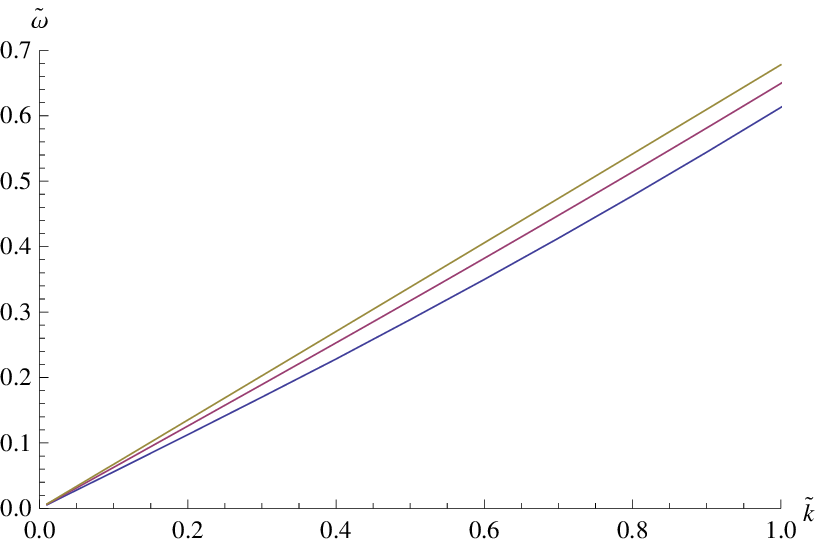}
\caption{(Left) The mass of the lowest two normal modes as a function of $n$.  We have chosen $\tilde m = -3$ and $\psi_\infty = \frac{1}{2} \arccos\left(\frac{1}{3}\right)$, which yields $n_{crit} = 0.565$.  The modes cross at $n_{cross} \approx -0.6$. (Right) The dispersion of the gapless mode at $n_{crit}$ for several values of $\tilde{m}$: The bottom blue curve is $\tilde m =  -3$, the middle red curve is $\tilde m =  -5$, and the top brown curve is $\tilde m = -20$.}
\label{T=0}
\end{figure}

At $n=0$ and zero temperature, the spectrum of normal modes is gapped \cite{Jokela:2010nu}, as expected in a quantum Hall state.  The first excited state is due to a pole in the current-current correlator, and the next mode comes from fluctuations of the embedding scalar $\tilde\rho$.  In Fig.~\ref{T=0}(L), we show the masses $\tilde\omega(\tilde k=0)$ of the first two excited states as a function of $n$.   The mass of the embedding scalar fluctuation does not depend on $n$ since it decouples from the fluctuations of the gauge fields at zero momentum.  The two lowest modes cross at $n_{cross}$.  As described in \cite{Jokela:2010nu, Jokela:2011sw}, near this level crossing, a magneto-roton develops in the spectrum at some nonzero momentum.  

\begin{figure}[ht]
\center
\includegraphics[width=0.475\textwidth]{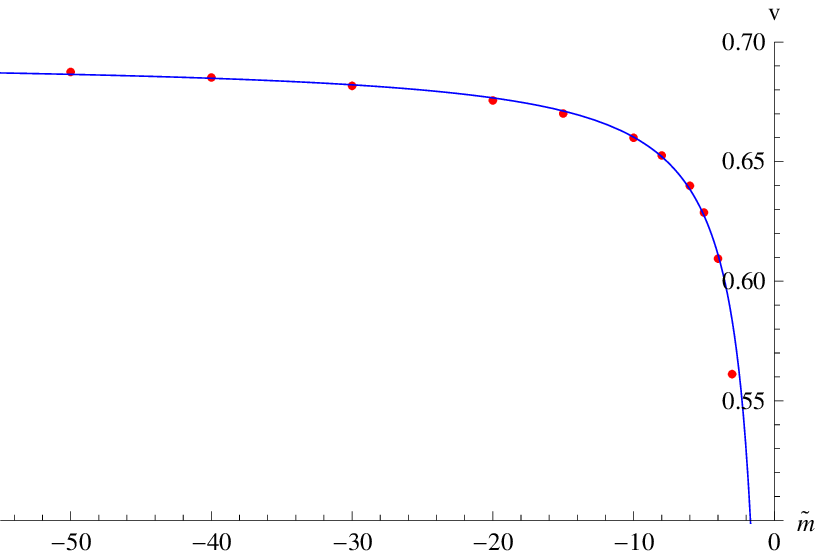}
\includegraphics[width=0.475\textwidth]{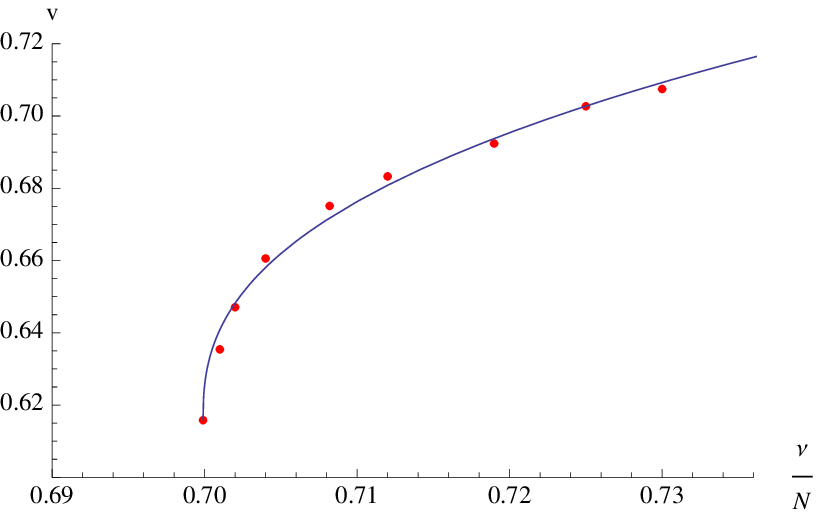}
\caption{(Left) The velocity of the gapless mode as a function of $\tilde{m}$.  The points are computed numerically and have been fit to the curve $v =  0.693+0.328/\tilde m$. (Right) The velocity of the massless mode as a function of the original filling fraction with $\tilde{m} = -20$ .  The curve to which the points are fit is $v = 0.62+0.38 (\nu/N-0.70)^{0.40}$.  In both plots $T=0$.}
\label{velocityT=0}
\end{figure}

As we approach $n_{crit}$, for which the transformed background magnetic field vanishes, the mass of the lowest mode decreases, and at $n_{crit}$ there is a gapless excitation, as expected from (\ref{ncomp}).  At this value of $n$, the system is a superfluid.  Figure~\ref{T=0}(R) shows the dispersion of this massless mode for different values of $\tilde{m}$. We plot in Fig.~\ref{velocityT=0}(L)  the velocity $v=\tilde\omega/\tilde k$, $\tilde k\to 0$, of the massless mode as a function of the mass parameter $\tilde{m}$.  %\footnote{The points are the slopes at $\tilde k=0$ of the curves in Fig.~\ref{T=0}(R).}
Figure~\ref{velocityT=0}(R) shows the velocity for different values of the original filling fraction $\nu$ at $n=0$, that is, for different kinds of anyons.

To compare with the results for free anyons, \cite{Chen:1989xs} gives the velocity of the gapless mode to be (in our notation) $v \sim \sqrt{\frac{\nu}{N}}\frac{\sqrt b}{m}$.
Note that this result is independent of whether we consider anyons to be made out of quarks and flux or baryons and flux.  As can be seen from Fig.~\ref{T=0}, we find that $v$ scales as $1/\tilde m$ and $\nu^{0.4}$ and with constant offsets. %Match this to our actual results

The  difference between the free anyon result  and our strong-coupling result
is not only in the actual form of the function but also in the meaning of the filling fraction $\nu$. To see this, note that the free anyon result predicts that the velocity of the massless mode changes when the filling fraction changes by one.\footnote{In fact, \cite{Chen:1989xs} only considered situations where $\nu$ is a large integer.}
In our case,  the equation of motion for the fluctuations depend only on $\psi_{\infty}$, and the boundary condition on $n$. Starting with some filling fraction $\nu$ and some boundary condition labelled by $n$, we can make a $T$ transformation, which changes $v \rightarrow \nu +1$ and leaves $n$ the same.  While $\nu$ has changed, the bulk fluctuations equation of motion has not, since the change came from a redefinition of the conserved current via boundary terms. This means that the normal mode spectrum will be unchanged.  Figure~\ref{velocityT=0}(R) is a result of changing $\psi_{\infty}$.
%I don't understand this paragraph, maybe put it in terms of \nu and \tilde\nu

\begin{figure}[ht]
\center
\includegraphics[width=0.475\textwidth]{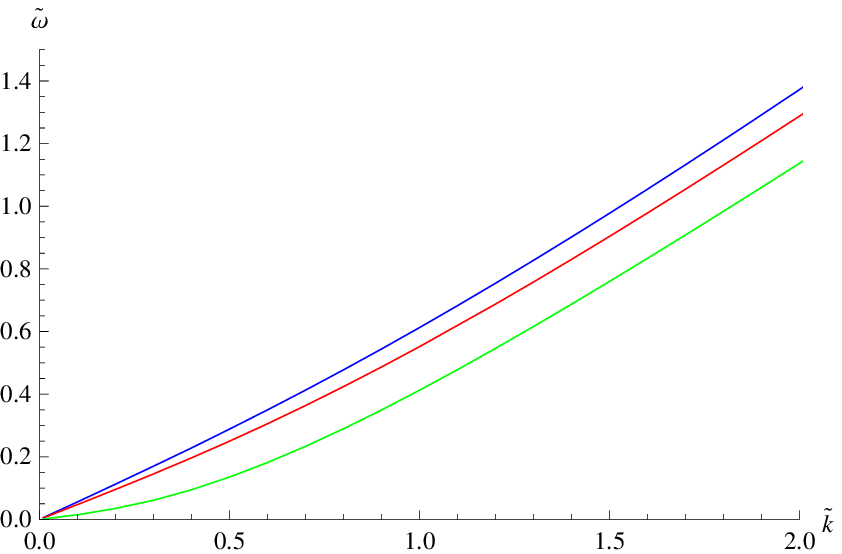}
\includegraphics[width=0.475\textwidth]{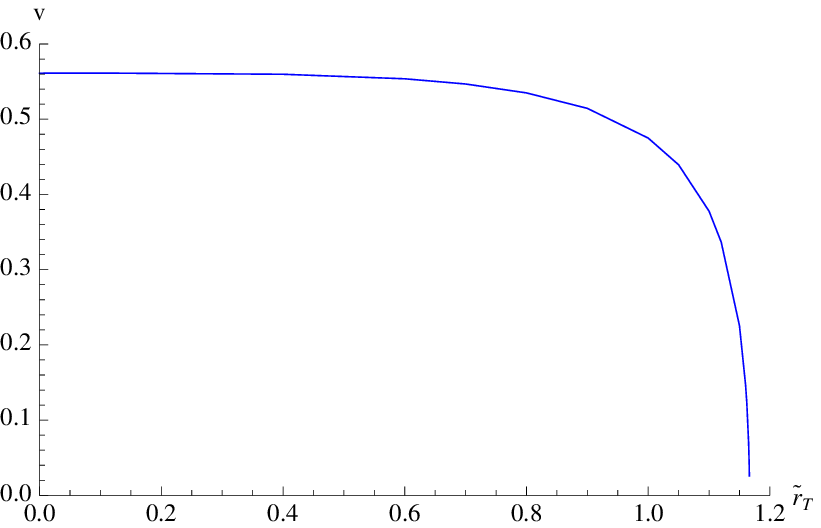}
\caption{(Left) The dispersion of the gaplesss mode with $\tilde m = -3$ for different temperatures:  The blue curve is $\tilde r_T = 0$, the red curve is $\tilde r_T = 1$,  and the green curve is $\tilde r_T = 1.16$. (Right) The velocity of the massless mode as a function of temperature, also with $\tilde m = -3$. At $\tilde r_T \approx 1.17$, the velocity vanishes, and the massless mode changes into a pair of purely imaginary mode, one of which has positive imaginary frequency, signaling an instability.}
\label{Tneq0}
\end{figure}

Turning on a small nonzero temperature does not qualitatively change the system.  In particular, there is still a gapless mode at $n_{crit}$ with an approximately linear dispersion, as shown in Fig.~\ref{Tneq0}(L).  This is in contrast to the nonzero-temperature behavior of regular holographic superfluids, where at nonzero temperature, the Goldstone mode frequency acquires an imaginary part \cite{Amado:2009ts, Herzog:2009ci}.  For a sufficiently high temperature, however, the massless mode velocity goes to zero, as can be seen in Fig.~\ref{Tneq0}(R), and an instability sets in.  Note that as described in \cite{Bergman:2010gm} (adapted to the current notation), as the temperature is increased, at around $\tilde{r}_{T}\approx 1.05$ two BH solutions start to exist, and at around $\tilde{r}_{T}\approx 1.23$, the MN embedding ceases to be a solution. Somewhere in between there is a first-order phase transition to the metallic BH phase, which is not a superfluid. 
 
\subsection{Comments}

Each value of $n$ in Fig.~\ref{T=0}(L) corresponds to an anyonic system in the presence of a residual magnetic field $B^*$.  As $n \rightarrow n_{crit}$, this residual magnetic field vanishes.  For any other $n$, we have an anyonic system which is gapped in the presence of the residual magnetic field.  This is a quantum Hall state in which the charge carriers are anyons with filling fraction $\nu^*$ depending on the $SL(2,{\mathbb Z})$ transformation (\ref{nuSL2Z}).  However, $n$ depends only on $c_s/d_s$ (\ref{ndef}) and not $a_s$ or $b_s$. At $n_{crit}$, there is a gapless mode, and the anyonic system is a superfluid. One can write solutions of the bulk equations of motion for a current flowing without any electric field driving it.
These are just a reinterpretation via the alternative quantization of the original solutions for the quantum Hall state with an electric field driving a Hall current.

We can ask what happens when we increase the magnetic field for fixed anyonic statistics, that is, for fixed $n$.  This can be done by starting with a different ratio of charge to magnetic field.  In this case, we are no longer in a MN embedding but rather a BH embedding. After changing the boundary conditions, the system is neither in a quantum Hall (for general $n$) or superfluid ($n=n_{crit}$) state.  Instead, it is in the metallic state represented by a BH embedding. 

 If we want to add just a finite number of magnetic fluxes, we can do that by starting with some extra amount of localized charge in the original theory. The charge is represented by elementary strings stretching from the tip of the D7-brane to the horizon. If there is a density of such strings, they will pull the D7-brane into the horizon.  But, if there are only a finite number of localized strings, they create local funnels connecting the D7-brane to the horizon.  In such places, locally $\sigma_{xx} \neq 0$.  After the appropriate $SL(2,{\mathbb Z})$ transformation, the strings will become magnetic vortices inside the superfluid.  However, at their cores, there is a normal fluid since originally $\sigma_{xx} \neq 0$ there.  
 If coupled to real, dynamical electromagnetism, the anyonic superfluid will therefore be a type II superconductor.
If the original quantum Hall system included impurities, there would be plateaux in $\sigma_{yx}$; we could add some extra charges, and the system would remain in the gapped MN phase.  The transformed system would then be stable to the introduction of some magnetic flux and would stay a superfluid even at some nonzero magnetic field.

\vspace{0.5cm}

{\bf \large Acknowledgments}
We especially thank  G. Semenoff for valuable suggestions which led to this work.  We also thank
O. Bergman, J. Sonnenschein, and A. Stern for discussions.
N.J. is funded in part by the Spanish grant FPA2011-22594, by Xunta de
Galicia (Conselleria de Educaci\'on, grant INCITE09-206-121-PR and grant PGIDIT10PXIB206075PR), by the Consolider-Ingenio 2010 Programme CPAN (CSD2007-00042), and by FEDER. N.J. is
also supported by the Juan de la Cierva program.  
The work of G.L was supported in part by the Israel Science Foundation under grant no.~392/09 and in part by a grant from the GIF, the German-Israeli Foundation for Scientific Research and Development under grant no. 1156-124.7/2011.
M.L. is also supported by funding from the European Research
Council under the European Union's Seventh Framework Programme (FP7/2007-2013) /
ERC Grant agreement no.~268088-EMERGRAV.  M.L. would like to thank the University of Santiago de Compostela for warm Galician hospitality.
%something here also about the Technion workshop

\appendix
\section{Fluctuation equations of motion}\label{appendix}

In this appendix we will provide the equations of motion for all the fluctuation fields studied in the bulk part of the paper. The equations
are easy and straightforward to obtain, albeit they are very long. To avoid extra clutter, we have omitted tildes in all the quantities.

The equation of motion for $\delta a_t$ is:
\bea
\label{atEOM}
 \partial_R H = -\frac{ik}{2}\delta a_y c'(R)-\frac{gk^2a'_0 P}{h^2r^6}\delta\rho+k\delta e_x\left(\frac{4G r^4}{g(1+r^4)}+\frac{g a'^2_0}{h^2r^4}\right)-ik\delta a_y\partial_R\left(\frac{g a'_0}{hr^4}\right) \ ,
\eea
where we introduced a fluctuation-dependent function $H$:
\bea
 H & = & \frac{16 R^3 \delta \rho \rho^2}{r^6}+\frac{g \left(1+r^4\right) \left(4 G h^2 r^4 R^4+g^2 \left(1+r^4\right) a_0'^2\right)\delta a_t'}{4 G h^3 r^8 R^4} \nonumber\\
 & & -\frac{g^3 P \left(1+r^4\right)^2 a_0' \delta \rho'}{4 G h^3 r^{10} R^4} \nonumber\\
 & & +\frac{1}{4 G h^3 r^{12} R^4}g \left(1+r^4\right) \delta\rho a_0' \Bigg(-\frac{16 G h^2 r^6 \left(2+r^4\right) R^4 \rho}{1+r^4} \nonumber\\
  && +8 h^2 r^4 R^4 \rho\left(f^2 R^2+\left(4+f^2\right) \rho^2\right) \nonumber\\
  && +g^2 \Big\{\rho \left(-h \left(r^2+r^6\right)+\left(5+3 r^4\right) R^2+\left(-2+7 h+(-2+5 h) r^4\right) \rho^2-2 r^2 \left(2+r^4\right) a_0'^2\right)\nonumber\\
  &&+(-1+h) R \left(r^2+r^6-2 \left(7+5 r^4\right) \rho^2\right) \rho' \nonumber\\
  &&-\rho \left(r^2+r^6+\left(2-7 h+2 r^4-5 h r^4\right) R^2-\left(5+3 r^4\right) \rho^2\right) \rho'^2\Big\}\Bigg) \ .
\eea
The equation of motion for $\delta a_x$ reads:
\bea
\label{axEOM}
 && \partial_R\left(\frac{g}{\omega}(k\delta a'_t-\delta e'_x)\right) \\
 & & = -i\omega\delta a_y\partial_R\left(\frac{g a'_0}{hr^4}\right)-\frac{gk\omega a'_0 P}{h^2r^6}\delta\rho+\frac{g\omega}{h^2r^4}(A+a'^2_0)\delta e_x-\frac{i\omega}{2}\delta a_y c'(R)\nonumber \ .
\eea
The equation of motion for $\delta a_y$ is as follows:
\bea
 -\partial_R(g a'_y) & = & \frac{i}{2}\delta e_x c'(R)+ i\delta e_x\partial_R\left(\frac{g a'_0}{h r^4}\right)-ik\delta\rho\partial_R\left(\frac{gP}{hr^6}\right) \nonumber\\
  & & +\delta a_y\left(\frac{g\omega^2 a'^2_0}{h^2r^4}-\frac{4Gr^4(hk^2r^4-(1+r^4)\omega^2)}{g(1+r^4)^2}\right)+i\frac{gkR\rho'}{hr^6}(1-h)\delta\rho \nonumber\\
  & & -16i\frac{kR^3\rho^2 a'_0}{r^6}\delta\rho+ik N\delta\rho \ .
\label{ayEOM}
\eea
Next we write down the $\delta\rho$ equation of motion,
\bea
&& -\partial_R\left(\frac{g^3(1+r^4)^3a'_0 P}{4Gh^3 r^{10} R^4}\delta a'_t\right)+\partial_R\left(\frac{g^3(1+r^4)^2P^2}{4G h^3 r^{12}R^4}\delta\rho'\right) \nonumber\\
&&  -\partial_R\left(\frac{g(1+r^4)}{hr^6}(hR^2+\rho^2)\delta\rho'\right)+\partial_R(S \delta\rho) \nonumber\\
& = & -\frac{gk a'_0P}{h^2 r^6}\delta e'_x+ik\delta a_y\partial_R\left(\frac{g P}{hr^6}\right) \nonumber\\
& & -\frac{g}{h^2r^8}\delta\rho\left(hk^2R^4-(1+R^4)\omega^2+(hk^2-\omega^2)\rho^2(2R^2+\rho^2)-k^2r^2(hR^2+\rho^2)a'^2_0  \right) \nonumber\\
& & -16\frac{R^3\rho^2}{r^6}(ik a'_0\delta a_y+\delta a'_t)-32\frac{R^3\rho a'_0}{r^8}\delta\rho(R^2-2\rho^2) \nonumber\\
& & -T\delta a'_t-kU\delta a_y-V\delta\rho-W\delta \rho' \ .
\label{rhoEOM}
\eea

The $\delta z$ equation of motion decouples from all the other fluctuations:
\be
\label{zEOM}
 -\partial_R\left(gr^2\sqrt{1+r^4}\delta z'\right) = -\frac{g A}{h r^2}\sqrt{1+r^4}k^2\delta z+\frac{g\sqrt{1+r^4}\omega^2}{h^2r^2}(A+a'^2_0)\delta z \ .
\ee
Finally, the constraint stemming from maintaining the radial gauge, $a_R=0$, reads:
\be
\label{aREOM}
 \frac{k}{\omega}g(\delta e'_x-k\delta a'_t) -\omega H= 0 \ .
\ee

The following background functions were introduced in the above equations,
\bea
 P = (1-h)R\rho+\rho'(hR^2+\rho^2)
\eea
and
\bea
 N & = & \frac{8 f^2 h R^6 \rho}{g r^4+g r^8}+\frac{g \left(h \left(r^2+r^6\right)-\left(5+7 r^4\right) R^2\right) \rho}{h r^8 \left(1+r^4\right)} \nonumber\\
   & & +\frac{\left(g^2 \left(2-7 h+2 r^4-9 h r^4\right)+8 \left(4+f^2\right) h^2 r^4 R^4\right) \rho^3}{g h r^8 \left(1+r^4\right)} \nonumber\\
   && +\frac{2 g \left(2+3 r^4\right) \rho a_0'^2}{h r^6 \left(1+r^4\right)}-\frac{g (-1+h) R \left(r^2+r^6-2 \left(7+9 r^4\right) \rho^2\right) \rho'}{h r^8 \left(1+r^4\right)} \nonumber\\
   & & +\frac{g \rho \left(r^2+r^6+\left(2-7 h+2 r^4-9 h r^4\right) R^2-\left(5+7 r^4\right) \rho^2\right) \rho'^2}{h r^8 \left(1+r^4\right)} \ ,
\eea
and
\bea
 S & = & \frac{g^3 P \left(1+r^4\right) \left(2+r^4\right) \rho a_0'^2}{2 G h^3 r^{12} R^4} -\frac{g (1-h) R \left(r^2+r^6-2 \left(7+5 r^4\right) \rho^2\right)}{h r^8} \nonumber\\
 & & +\frac{g^3 P \left(1+r^4\right) \rho\left(h \left(r^2+r^6\right)-\left(5+3 r^4\right) R^2+\left(2-7 h+2 r^4-5 h r^4\right) \rho^2\right)}{4 G h^3 r^{14} R^4} \nonumber\\
 & & -\frac{2 g P \left(1+r^4\right) \rho\left(f^2 R^2+\left(4+f^2\right) \rho^2\right)}{G h r^{10}} \nonumber\\
 & & -\frac{2 g \rho\left(r^2+r^6+\left(2-7 h+2 r^4-5 h r^4\right) R^2-\left(5+3 r^4\right) \rho^2\right) \rho'}{h r^8} \nonumber\\
 & & -\frac{g^3 (-1+h) P \left(1+r^4\right) \left(r^2+r^6-2 \left(7+5 r^4\right) \rho^2\right) \rho'}{4 G h^3 r^{14} R^3} \nonumber\\ 
 & & +\frac{g^3 P \left(1+r^4\right) \rho\left(r^2+r^6+\left(2-7 h+2 r^4-5 h r^4\right) R^2-\left(5+3 r^4\right) \rho^2\right) \rho'^2}{4 G h^3 r^{14} R^4} \ ,
\eea
and
\bea
 T & = &  -\frac{g^3 \left(-1-r^4\right) \left(2+r^4\right) \rho a_0'^3}{2 G h^3 r^{10} R^4} \nonumber\\
 && +g^3 \frac{\left(1+r^4\right) \rho\left(h \left(r^2+r^6\right)-\left(5+3 r^4\right) R^2+\left(2-7 h+2 r^4-5 h r^4\right) \rho^2\right)}{4 G h^3 r^{12} R^4}a_0' \nonumber\\
 && -\frac{2 g \rho\left(-2 G r^2 \left(2+r^4\right)+f^2 \left(1+r^4\right) R^2+\left(4+f^2\right) \left(1+r^4\right) \rho^2\right) a_0'}{G h r^8} \nonumber\\
 && +\frac{g^3 (-1+h) \left(-1-r^4\right) \left(r^2+r^6-2 \left(7+5 r^4\right) \rho^2\right) a_0' \rho'}{4 G h^3 r^{12} R^3} \nonumber\\
 && -\frac{g^3 \left(-1-r^4\right) \rho\left(r^2+r^6+\left(2-7 h+2 r^4-5 h r^4\right) R^2-\left(5+3 r^4\right) \rho^2\right) a_0' \rho'^2}{4 G h^3 r^{12} R^4} \ ,
\eea
and
\bea
 U & = & \frac{8 i f^2 h R^6 \rho}{g r^4(1+r^4)}+\frac{i g \left(h \left(r^2+r^6\right)-\left(5+7 r^4\right) R^2\right) \rho}{h r^8 \left(1+r^4\right)} \nonumber\\
  &  & +\frac{i \left(g^2 \left(2-7 h+2 r^4-9 h r^4\right)+8 \left(4+f^2\right) h^2 r^4 R^4\right) \rho^3}{g h r^8 \left(1+r^4\right)}+\frac{2 i g \left(2+3 r^4\right) \rho a_0'^2}{h r^6 \left(1+r^4\right)} \nonumber\\
   & &-\frac{i g (-1+h) R \left(r^2+r^6-2 \left(7+9 r^4\right) \rho^2\right) \rho'}{h r^8 \left(1+r^4\right)} \nonumber\\
   & & +\frac{i g \rho\left(r^2+r^6+\left(2-7 h+2 r^4-9 h r^4\right) R^2-\left(5+7 r^4\right) \rho^2\right) \rho'^2}{h r^8 \left(1+r^4\right)} \ ,
\eea
and
\bea
 W & = & -\frac{g^3 P \left(1+r^4\right) \left(2+r^4\right) \rho a_0'^2}{2 G h^3 r^{12} R^4} \nonumber\\
  & & -\frac{g^3 P \left(1+r^4\right) \rho \left(r^2+r^6+\left(2-7 h+2 r^4-5 h r^4\right) R^2-\left(5+3 r^4\right) \rho^2\right) \rho'^2}{4 G h^3 r^{14} R^4} \nonumber\\
  && +\frac{g (1-h) R \left(r^2+r^6-2 \left(7+5 r^4\right) \rho^2\right)}{h r^8} \nonumber\\
  && +\frac{2 g \rho \left(r^2+r^6+\left(2-7 h+2 r^4-5 h r^4\right) R^2-\left(5+3 r^4\right) \rho^2\right) \rho'}{h r^8} \nonumber\\
  && +\frac{1}{4 G h^3 r^{14} R^4}g P \left(1+r^4\right) \Bigg(\rho g^2 \left(-h \left(r^2+r^6\right)+\left(5+3 r^4\right) R^2\right)\nonumber\\
  &&+\left(g^2 \left(-2+7 h+(-2+5 h) r^4\right)+8 \left(4+f^2\right) h^2 r^4 R^4\right) \rho^3\nonumber\\
  &&+8\rho f^2 h^2 r^4 R^6+g^2 (-1+h) R \left(r^2+r^6-2 \left(7+5 r^4\right) \rho^2\right) \rho'\Bigg) \ ,
\eea
and finally
\bea
 V & = & -\frac{g^3 \left(2+r^4\right)^2 \rho^2 a_0'^4}{G h^3 r^{12} R^4} \nonumber\\
 && +g\frac{8 \left(2+r^4\right) \rho^2 \left(f^2 R^2+\left(4+f^2\right) \rho^2\right) a_0'^2}{G h r^{10}} \nonumber\\
 && +g\frac{\left(8 h^2 r^{10} \left(2+r^4\right) R^4-16 h^2 r^8 \left(10+3 r^4\right) R^4 \rho^2\right) a_0'^2}{4 h^3 r^{16} R^4} \nonumber\\
 && -\frac{1}{G h^3 r^{14} R^4}g^3 \left(2+r^4\right) \rho a_0'^2\Bigg(-(-1+h) r^2 \left(1+r^4\right) R \rho' \nonumber\\
 && +2 (-1+h) \left(7+5 r^4\right) R \rho^2 \rho'-\rho^3 \left(-2 \left(1+r^4\right)+h \left(7+5 r^4\right)+\left(5+3 r^4\right) \rho'^2\right)\nonumber\\
 && +\rho \left(h \left(r^2+r^6\right)-\left(5+3 r^4\right) R^2 +\left(r^2+r^6+(2-7 h) R^2+(2-5 h) r^4 R^2\right) \rho'^2\right)\Bigg) \nonumber\\
 & & +\frac{8 h R^4 \left(f^4 R^6+\left(4+f^2\right) \rho^2 \left(3 f^2 R^4+3 f^2 R^2 \rho^2+\left(4+f^2\right) \rho^4\right)\right)}{g G r^8} \nonumber\\
 &&+\frac{1}{h r^{10}}g \Bigg(h \left(r^4+r^8\right)-r^2 \left(5+3 r^4\right) R^2\nonumber\\
 &&+\rho^2 \Big\{r^2 \left(8-33 h+8 r^4-23 h r^4\right)-2 \left(-31+h+(-13+h) r^4\right) R^2\nonumber\\
 &&+2 \left(-25-17 r^4+h \left(55+29 r^4\right)\right) \rho^2\Big\}\nonumber\\
&& +2 (-1+h) R \rho\left(r^2 \left(19+13 r^4\right)+2 \left(1+r^4\right) R^2-2 \left(55+29 r^4\right) \rho^2\right) \rho'\nonumber\\
&&+\Big\{r^4+r^8-h r^2 \left(5+3 r^4\right) R^2-2 (-1+h) \left(1+r^4\right) R^4 \nonumber\\
&&+\rho^2 \left(-5 r^2 \left(5+3 r^4\right)+2 \left(-25-17 r^4+h \left(55+29 r^4\right)\right) R^2 +12 \left(5+2 r^4\right) \rho^2\right)\Big\} \rho'^2\Bigg) \nonumber\\
&& -\frac{1}{4 G h^3 r^{16} R^4}g \Bigg\{\rho\left(h \left(r^2+r^6\right)-\left(5+3 r^4\right) R^2+\left(2-7 h+2 r^4-5 h r^4\right) \rho^2\right) \nonumber\\
&& -(-1+h) R \left(r^2+r^6-2 \left(7+5 r^4\right) \rho^2\right) \rho' \nonumber\\
&&+\rho\left(r^2+r^6+\left(2-7 h+2 r^4-5 h r^4\right) R^2-\left(5+3 r^4\right) \rho^2\right) \rho'^2\Bigg\}\nonumber\\
&&\times\Bigg\{-16 \rho f^2 h^2 r^4 R^6+16\rho \left(4+f^2\right) h^2 r^4 R^4 \nonumber\\
&& +\rho \left(g^2 \left(h \left(r^2+r^6\right)-\left(5+3 r^4\right) R^2\right)-\left(g^2 \left(-2+7 h+(-2+5 h) r^4\right)\right)\right) \rho^2 \nonumber\\
&& -g^2 (-1+h) R \left(r^2+r^6-2 \left(7+5 r^4\right) \rho^2\right) \rho' \nonumber\\
&&+g^2 \rho \left(r^2+r^6+\left(2-7 h+2 r^4-5 h r^4\right) R^2-\left(5+3 r^4\right) \rho^2\right)\rho'^2\Bigg\} \ .
\eea

\end{document}